\documentclass[final,3p,times]{elsarticle}

\usepackage{graphicx}
\usepackage{algorithm}          
\usepackage{algpseudocode}   
\usepackage{tikz}

\usepackage{booktabs}
\usepackage{amssymb}
\usepackage{epstopdf}
\usepackage{lmodern}
\usepackage{amsmath, amsthm}
\usepackage[T1]{fontenc}
\usepackage{color}
\usepackage{float}
\usepackage{tabularx}
\usepackage{subcaption}
\usepackage{pgfplots}
\pgfplotsset{compat=1.5}

\biboptions{numbers,sort&compress} 

\journal{JOURNAL NAME}

\begin{document}

\begin{frontmatter}

\title{A Stabilized Diffuse-Interface Electroporation Model with a Semi-Analytical Spectral Electrolyte Solver}

\author[bu]{Saman Seifi}
\ead{samansei@bu.edu}

\address[bu]{Department of Mechanical Engineering, Boston University, Boston, MA 02215, USA}

\author[ub]{David Salac}
\ead{davidsal@buffalo.edu}
\address[ub]{Department of Mechanical and Aerospace Engineering, University at Buffalo--SUNY, Buffalo, NY 14260, USA}

\begin{abstract}
We develop a diffuse-interface continuum model for membrane electroporation that couples a phase field for pore geometry to a quasi-static electrolyte potential and a spatially varying leaky-dielectric model for the transmembrane voltage. The main contribution is a stabilized time-integration strategy for transmembrane voltage $V_m$: the stiff leakage term is treated implicitly while the electrolyte-to-membrane ionic current is lagged, yielding a closed-form update that removes the restriction imposed by the fast dielectric relaxation time. The electrolyte potential is computed efficiently using a semi-analytical spectral Laplace solver: a 2D DCT in the membrane plane reduces the 3D problem to independent 1D ODEs in $z$, solved in closed form and reconstructed by an inverse transform. The coupled method is robust under grid refinement, reproduces the sharp-interface critical-radius bifurcation, and captures electric-field focusing through conductive pores. We also demonstrate stochastic pore nucleation by adding thermal noise to the phase-field dynamics, enabling fully emergent electroporation events without prescribing initial defects.
\end{abstract}


\end{frontmatter}

\section{Introduction}
Synthetic lipid vesicles (e.g., liposomes) have emerged as versatile carriers for drug and gene delivery in biomedical applications. These structures can encapsulate therapeutic molecules such as drugs or nucleic acids and deliver them to specific tissues or cells, enhancing treatment efficacy while minimizing systemic toxicity \citep{Allen2004,Torchilin2005,Basak2025,Sharma2024}. A major challenge in vesicle-based therapeutics, however, lies in achieving precise spatiotemporal control over the release of encapsulated agents.

One promising strategy for controlled release is electroporation, the use of high electric fields to transiently permeabilize lipid membranes by inducing pore formation \citep{Neumann1989,Tsong1991,Weaver1996}. This mechanism offers a non-viral alternative for delivering macromolecules into cells and has shown potential in gene therapy, immunotherapy, and tumor treatment \citep{Gehl2003,Yarmush2014,Rakoczy2022,Campelo2023}. While adoption varies by indication and geography, several electroporation-based interventions are now used clinically, including electrochemotherapy (ECT) for cutaneous and subcutaneous tumors under ESOPE guidelines, FDA-approved pulsed field ablation (PFA) for atrial fibrillation, and FDA-cleared irreversible electroporation (IRE) for prostate tissue ablation \citep{Marty2006,Mir2006,FDAPFA2024,AngioIRE2024}. A key limiting factor remains the absence of comprehensive, predictive models that capture the multi-scale physics of membrane poration \citep{Campelo2023}.

Electroporation is now widely accepted to proceed via electrical breakdown of lipid membranes, characterized by the formation of transient hydrophilic pores at the nanometer scale \citep{Weaver1996,Smith2004,DeBruin1999}. The application of a strong transmembrane potential increases the membrane surface tension and reduces the energy barrier for defect formation; once a critical threshold is exceeded, local membrane regions destabilize, leading to nucleation and pore expansion \citep{Boeckmann2008,Kasparyan2024}. Simple energetic or 1D pore models capture the line-tension vs.\ membrane surface tension tradeoff but typically omit spatial electromechanics and thus have limited quantitative fidelity \citep{MOROZ19972211,ZHELEV199389,karatekin2003cascades}. Atomistic simulations have revealed detailed lipid rearrangements, water intrusion, and pore stabilization under electric stress \citep{LEONTIADOU20042156,TAREK20054045,Boeckmann2008}, but such methods remain computationally limited in size and timescale.  Continuum formulations can bridge these regimes; however, many have treated only portions of the coupling e.g., solving for a uniform transmembrane voltage or decoupling electrolyte transport from membrane dynamics or ignoring electrical couplings altogether~\citep{DeBruin1999,Aubin_Ryham_2016,jaramillo2023phase,seifi2016phasefieldmodelingelectricfield}. A comprehensive mathematical formulation and efficient numerical method that solves from electrodes (electrostatics) through membrane charging to pore evolution remains underdeveloped.

We address this gap with a diffuse-interface (phase-field) framework  that couples Allen-Cahn~\citep{Allen1979} pore evolution to a three-dimensional semi-analytical Laplace solve for the electrolyte and a spatially varying leaky-dielectric model for the transmembrane voltage. The scheme is stabilized by a semi-implicit treatment that handles the stiffness induced by disparate time scales (fast dielectric relaxation vs.\ slower charging and pore dynamics), yielding efficient and robust computations~\citep{CHEN1998147,KIM20062554,PhysRevLett.94.146103} . The model reproduces the critical radius bifurcation and captures electric-field focusing, two hallmarks of electroporation multi-physics, providing a compact, predictive continuum tool for electrically mediated membrane permeabilization.

By simulating pore dynamics with a phase-field model, this work advances the theoretical basis of controlled membrane permeabilization and provides design tools for tunable release protocols in biomedical applications.

\section{Continuum Model Formulation}
\subsection{Diffuse Interface Model of Lipid Membrane}
The competition between pore edge energy a.k.a line tension \(\gamma\) with units of \(\mathrm{J\,m^{-1}}\) and effective membrane surface tension \(\sigma_{\mathrm{eff}}\) with unit of \(\mathrm{J\,m^{-2}}\) drives pore dynamics. Line tension promotes closure because the rim is energetically costly: lipids at the edge lose favorable tail-tail contacts, must reorient to shield hydrophobes, and concentrate local bending; minimizing rim length lowers this penalty. In contrast, surface tension promotes opening because creating a hole relieves lateral stretch, reducing the elastic cost associated with elevated area per lipid and releasing stored energy. In this work we adopt a \emph{flat-membrane} model: Helfrich curvature terms are not included explicitly (mean curvature \(H\!\approx\!0\) away from the rim), and curvature/topology effects localized at the rim are absorbed into an effective line tension~\citep{Helfrich1973}.

\begin{figure}[ht!]
\centering
\begin{subfigure}[c]{0.27\textwidth}
    \includegraphics[width=\textwidth]{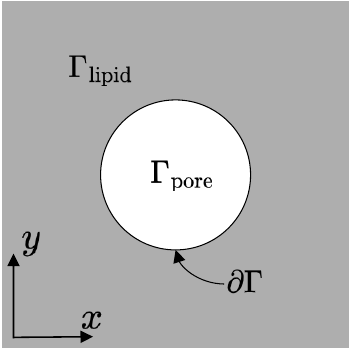}
    \caption{}
\end{subfigure}%
\hspace{0.02\textwidth} 
\begin{subfigure}[c]{0.5\textwidth}
    \includegraphics[width=\textwidth]{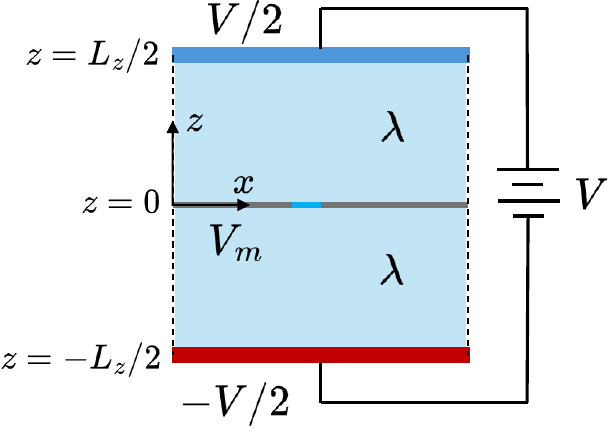}
    \caption{}
\end{subfigure}
\caption{(a) The computational domain $\Gamma = \Gamma_{\text{lipid}} \cup \Gamma_{\text{pore}}$. (b) A cross section of the planar lipid membrane placed between parallel electrodes that impose a uniform potential
difference $V$}
\label{fig:domains}
\end{figure}

For a flat membrane domain \(\Gamma\subset\mathbb{R}^{2}\) with pore region \(\Gamma_{\mathrm{pore}}\), lipid region \(\Gamma_{\mathrm{lipid}}\), and the lipid/pore boundary \(\partial\Gamma\) (Fig.~\ref{fig:domains}a), the sharp-interface free energy is \citep{deryagin1962theory,tu2003lipid,capovilla2002lipid}
\begin{equation}
E
\;=\;
\oint_{\partial\Gamma} \gamma\, ds
\;+\;
\int_{\Gamma_{\mathrm{lipid}}} \sigma_{\mathrm{eff}}\, dA,
\label{eq:sharp_interface}
\end{equation}
where the first term penalizes rim length and the second term represents the energetic cost of the lipid region, driving the expansion of the pore area.

We then adopt a diffuse-interface approach by introducing a phase-field variable $\phi(\mathbf{r}(x,y), t)$, smoothly varying between $\phi=1$ where it represents the lipid region and $\phi=0$ for the aqueous pore in $\Gamma \subset \mathbb{R}^2$. The Ginzburg-Landau free energy functional by assuming the isotropic line tension is adopted to approximate the energy functional $E$ can be written as:
\begin{equation}
E[\phi] = \frac{\gamma}{C_g}\int_{\Gamma}\!\left(\frac{\varepsilon}{2}\,|\nabla\phi|^{2} + \frac{g(\phi)}{\varepsilon}\right)\,dA + \int_{\Gamma}\!\sigma_{\text{eff}}H(\phi)\,dA,
\label{eq:E_pf}
\end{equation}
where $\varepsilon$ is the interface thickness, $g(\phi) = \frac{1}{4}\phi^2(1-\phi)^2$ is a double-well potential, and $H(\phi) = \phi^2(3-2\phi)$ is a smooth interpolation function. The constant $C_g=\sqrt{2}\int_{0}^{1}\!\sqrt{g(\phi)}\,d\phi=\sqrt{2}/12$ ensures that the model recovers the correct line tension $\gamma$ in the sharp-interface limit.

This functional is the continuum, diffuse-interface equivalent of the classical hydrophilic pore model \citep{ABIROR197937}, where $\gamma$ is a constant, effective line tension. 
In this work we focus on the numerical stabilization of the coupled electro-phase-field system, we adopt the standard and widely used Ginzburg-Landau formulation, where the complex physics of nucleation and reorientation are absorbed into the constant phenomenological line tension $\gamma$.

The evolution of the non-conserved order parameter $\phi$ follows an Allen-Cahn equation, representing local lipid rearrangement without global mass conservation on the timescale of pore formation:
\begin{equation}
\frac{\partial\phi}{\partial t} = -M \frac{\delta E}{\delta\phi} = -M \left[ \frac{\gamma}{C_g}\left(-\varepsilon\nabla^{2}\phi+\frac{g'(\phi)}{\varepsilon}\right) + \sigma_{\text{eff}} H'(\phi) \right],
\label{eq:AC_code}
\end{equation}
where $M$ is the phase-field mobility, a kinetic coefficient that sets the characteristic time scale for the local relaxation of the order parameter. This mobility is physically related to the dissipative mechanisms at the interface, which are governed by an effective viscosity $\eta_c$ at the pore edge ($M \propto 1/\eta_c$).
\subsection{Electromechanical Coupling}
We model electromechanical coupling through an effective lateral pressure arising from the Maxwell stress acting within the membrane plane. A transmembrane voltage $V_m(x,y,t)$ induces a compressive stress in the dielectric lipid core, which can be represented by the associated stored electrostatic energy density 
\begin{equation}
    f_{\text{elec}} = \frac{1}{2} C_{\text{lipid}} V_m^2.    
\end{equation}
In coupled diffuse-interface formulations, a well-known numerical artifact arises because the locally resolved $V_m$ collapses toward zero within the conductive pore region. This collapse occurs precisely at the pore rim where the phase-field driving force $H'(\phi)$ is maximal, which would artificially suppress the electrical contribution to the pore dynamics. To address this while preserving the underlying Maxwell stress physics, we treat the electrical contribution as a component of the membrane's far-field surface tension. We define the electrical pressure $P_{\text{elec}}$ as a non-local, spatially uniform quantity obtained by averaging the local electrostatic energy density over the intact lipid region:
\begin{equation}
    P_{\text{elec}} = \frac{\int_{\Gamma} f_{\text{elec}} H(\phi) \, dA}{\int_{\Gamma} H(\phi) \, dA}.
    \label{eq:p_elec_maxwell}
\end{equation}
This formulation effectively serves as a mean-field approximation, assuming that the relaxation of lateral tension within the lipid bilayer is significantly faster than the timescale of pore expansion. By utilizing this non-local pressure, we ensure that the total electrical energy stored in the membrane correctly drives the expansion of the pore rim, consistent with the observed behavior of tension-mediated pore growth in experimental lipid systems. The resulting pressure $P_{\text{elec}}$ is incorporated into the Allen--Cahn evolution as an addition to the base tension: 
\begin{equation}
    \sigma_{\text{eff}} = \sigma_0 + P_{\text{elec}} .
    \label{eq:effective_surface}
\end{equation}
\subsection{Electrostatic and Leaky Dielectric Model}
The transmembrane voltage $V_m(x,y,t)$ is determined via a coupled electrostatic model. In the 3D electrolyte, the electric potential $\Phi(x,y,z,t)$ is governed by the Laplace equation:
\begin{equation}
    \nabla^2\Phi = 0, 
\end{equation}
subject to an applied potential $\pm V/2$ at the electrodes located at the top and bottom boundaries ($z = \pm L_z/2$). 

We treat the membrane as a zero-thickness interface located at the midplane $z=0$, which partitions the computational domain into two electrolyte regions: the upper region with potential $\Phi^+(x,y,z,t)$ for $z>0$ and the lower region with potential $\Phi^-(x,y,z,t)$ for $z<0$. At this interface, we impose a potential jump corresponding to the transmembrane voltage and enforce continuity of the normal ionic current:
\begin{align}
    \Phi^{+}(x,y,0^+) - \Phi^{-}(x,y,0^-) &= V_m(x,y,t), \label{eq:phi_jump} \\
    \lambda^{+}\,\partial_z \Phi^{+}(x,y,0^+) &= \lambda^{-}\,\partial_z \Phi^{-}(x,y,0^-). \label{eq:flux_continuity}
\end{align}
Here, $\partial_z$ denotes differentiation in the electrode-normal direction, and $\lambda^{\pm}$ are the electrolyte conductivities in the upper and lower regions, respectively. Although the formulation allows for asymmetric electrolytes, we assume $\lambda^{+}=\lambda^{-}=\lambda$ in this work for simplicity.
Equation~\eqref{eq:phi_jump} represents the discontinuous potential drop across the lipid bilayer, while Eq.~\eqref{eq:flux_continuity} enforces conservation of ionic current across the membrane interface.
The temporal evolution of the transmembrane voltage $V_m(x,y,t)$ is governed by a spatially varying leaky-dielectric model,
\begin{equation}
 \frac{\partial V_m}{\partial t}
 =
 \frac{1}{C_m(\phi)}
 \left[
 J_{\mathrm{elec}} - G_m(\phi)\,V_m
 \right],
 \label{eq:governing_vm}
\end{equation}
where $C_m(\phi)$ and $G_m(\phi)$ denote the effective membrane capacitance and conductance, respectively.
The source term $J_{\mathrm{elec}}(x,y,t)$ represents the ionic current density supplied by the surrounding electrolyte to the membrane.
Consistent with the flux continuity condition \eqref{eq:flux_continuity}, we define
\begin{equation}
 J_{\mathrm{elec}}(x,y,t)
 \;\equiv\;
 \lambda\,\partial_z \Phi^{+}(x,y,0^+,t)
 \;=\;
 \lambda\,\partial_z \Phi^{-}(x,y,0^-,t),
 \label{eq:J_elec_def}
\end{equation}
where the equality follows from current conservation at the interface.
By this convention, $J_{\mathrm{elec}}>0$ corresponds to ionic current flowing from the bulk electrolyte into the membrane, contributing to capacitive charging or leakage through conductive pores.

The effective membrane capacitance $C_m(\phi)$ and conductance $G_m(\phi)$ are modeled as interpolations between the properties of the intact lipid bilayer and those of an aqueous pore, mediated by the phase-field variable $\phi$ through the interpolation function $H(\phi)$:
\begin{align}
    G_m(\phi) &= G_{\text{pore}} + \bigl(G_{\text{lipid}} - G_{\text{pore}}\bigr)H(\phi), \\
    C_m(\phi) &= C_{\text{pore}} + \bigl(C_{\text{lipid}} - C_{\text{pore}}\bigr)H(\phi).
\end{align}
The pore is treated as a highly conductive aqueous pathway with an effective conductance $G_{\text{pore}} = \lambda / d_m$, where $d_m$ represents the physical thickness of the membrane. Unlike discrete thickness models, this boundary condition approach allows the membrane properties to be defined as interface characteristics independent of the bulk electrolyte discretization.

\section{Numerical Method}
\subsection{Timescale Analysis and Mobility Selection}

The primary numerical challenge in the coupled electro--mechanical model arises from the strong separation of physical timescales. The dielectric relaxation (Debye) time of the electrolyte,
\[
\tau_{\mathrm d} = \frac{\epsilon_{\mathrm w}}{\lambda},
\]
is significantly shorter than the characteristic membrane charging time,
\[
\tau_{\mathrm{elec}} \sim \frac{L_z C_{\text{lipid}}}{2\lambda}.
\]
This disparity, $\tau_{\mathrm d} \ll \tau_{\mathrm{elec}}$, introduces substantial numerical stiffness into the governing equations.

A third relevant timescale is associated with pore evolution, $\tau_{\mathrm{phase}}$, governed by the Allen--Cahn phase-field mobility $M$. For a roughly circular pore of characteristic radius $R$, whose dynamics are driven by the line tension $\gamma$ of the pore rim, curvature-driven motion yields the scaling
\[
\tau_{\mathrm{phase}} \sim \frac{R^2}{M\gamma}.
\]
To accurately capture the interplay between electrical charging and membrane deformation, we choose the mobility such that pore evolution occurs on a comparable timescale to membrane charging. Enforcing the condition $\tau_{\mathrm{phase}} \approx \tau_{\mathrm{elec}}$ leads to the following physically motivated estimate for the mobility:
\begin{equation}
    M \sim \frac{2\lambda R^2}{\gamma L_z C_{\text{lipid}}}.
    \label{eq:mobility_scaling}
\end{equation}
This choice ensures that the phase-field evolution is neither artificially instantaneous nor excessively delayed relative to the electrical dynamics, allowing the semi-implicit numerical scheme to focus on the essential electromechanical coupling while maintaining stability.

\subsection{Phase-Field Solver}
Equation \eqref{eq:AC_code} is solved using a semi-implicit Fourier spectral method and imposing periodic boundary conditions in $x$-$y$ plane where the membrane is being modeled. The stiff Laplacian term ($\nabla^2\phi$) is treated implicitly for stability, while the non-linear term ($g'(\phi)$) is treated explicitly. In Fourier space, the update scheme for the transformed variable $\hat{\phi}$ is:
\begin{equation}
\hat{\phi}^{\,n+1}
=
\frac{
\hat{\phi}^{\,n}
+
\Delta t\,\mathcal{F}\!\left[
-M\left(
\frac{\gamma}{C_g}\frac{g'(\phi^n)}{\varepsilon}
+\sigma_{\mathrm{eff}}^{\,n}\,H'(\phi^n)
\right)
\right]
}{
1+\Delta t\,M\left(\frac{\gamma}{C_g}\right)\varepsilon\,k^2
}.
\label{eq:phi_hat}
\end{equation}

where $k^2 = k_x^2 + k_y^2$ is the squared wavenumber, $\mathcal{F}[\cdot]$ is the Fourier transform operator, and $\Delta t$ is the time step.
\subsection{Electrostatic solver}

The electric potential in the electrolyte, $\Phi(x,y,z,t)$, satisfies the quasi-static Laplace equation $\nabla^2\Phi=0$ in the rectangular domain $[0,L_x]\times[0,L_y]\times[-L_z/2,L_z/2]$. Homogeneous Neumann boundary conditions are imposed on the lateral boundaries $\partial\Phi/\partial n=0$ at $x=\{0,L_x\}$ and $y=\{0,L_y\}$. The top and bottom electrodes at $z=\pm L_z/2$ are held at fixed potentials $\Phi=\pm V_{\text{applied}}/2$, respectively.

\subsubsection{Singular interface formulation}
Rather than resolving the membrane as a volumetric region or a discrete grid gap, we treat the lipid bilayer as a singular interface at the midplane $z=0$. This interface partitions the domain into two electrolyte subregions: Region 1 ($-L_z/2 \le z < 0$) and Region 2 ($0 < z \le L_z/2$). The coupling between these regions is governed by the transmembrane voltage $V_m(x,y,t)$ through a potential jump condition and a requirement of current (flux) continuity:
\begin{align}
    \Phi(x,y,0^+,t) - \Phi(x,y,0^-,t) &= V_m(x,y,t), \label{eq:jump_jump} \\
    \lambda^+ \frac{\partial \Phi}{\partial z} \bigg|_{z=0^+} &= \lambda^- \frac{\partial \Phi}{\partial z} \bigg|_{z=0^-}, \label{eq:jump_flux}
\end{align}
where $0^\pm$ denote the limits approaching the interface from the top and bottom electrolyte regions, and $\lambda$ is the electrolyte conductivity. Equation~\eqref{eq:jump_flux} ensures that the ionic current density is conserved as it crosses the interface to charge the membrane capacitance or leak through pores.

\subsubsection{Spectral solution and interface coupling}
To exploit the lateral Neumann conditions, a two-dimensional discrete cosine transform (DCT-II) is applied to $\Phi$ on each $z$-slice. In spectral space, each transverse mode $(k_x,k_y)$ satisfies an uncoupled 1D Helmholtz equation:
\begin{equation}
\frac{d^2 \hat{\Phi}}{dz^2}-k_{xy}^2\hat{\Phi}=0, \qquad k_{xy}^2= \left(\frac{\pi k_x}{L_x}\right)^2+ \left(\frac{\pi k_y}{L_y}\right)^2.
\label{eq:hemlholtz_eq}
\end{equation}
The general solution for a mode in an electrolyte region of length $L = L_z/2$ is a linear combination of hyperbolic functions. By enforcing the boundary conditions at the electrodes ($\hat{\Phi}^{+}_{\mathrm{el}}, \hat{\Phi}^{-}_{\mathrm{el}}$) and the interface conditions in Eqs.~\eqref{eq:jump_jump}--\eqref{eq:jump_flux}, we solve for the unique interface potentials $\hat{\Phi}(0^\pm)$. For $k_{xy} > 0$:
\begin{equation}
    \hat{\Phi}(0^\pm) = \pm \frac{1}{2}\hat{V}_m + \frac{\hat{\Phi}_{\mathrm{el}}^{+} + \hat{\Phi}_{\mathrm{el}}^{-}}{2 \cosh(k_{xy} L)}.
    \label{eq:Phi_hat_bcs}
\end{equation}

Within each electrolyte region, the potential is then reconstructed analytically:
For each transverse spectral mode $(k_x,k_y)$ with $k_{xy}>0$, the electrolyte potential in either half-domain is reconstructed analytically as
\begin{equation}
\hat{\Phi}(z)=
\hat{\Phi}_{\mathrm{el}}^{\pm}
\frac{\sinh\!\big(k_{xy}(L-|z|)\big)}{\sinh\!\big(k_{xy}L\big)}
+
\hat{\Phi}(0^\pm)
\frac{\sinh\!\big(k_{xy}|z|\big)}{\sinh\!\big(k_{xy}L\big)},
\qquad
z \gtrless 0,
\label{eq:Phi_analytic}
\end{equation}
Where in Eq.~\eqref{eq:Phi_hat_bcs}, the quantities $\hat{\Phi}^{+}_{\mathrm{el}}$ and $\hat{\Phi}^{-}_{\mathrm{el}}$ denote the two-dimensional discrete cosine transform (DCT) coefficients of the electrode potentials imposed at the top and bottom boundaries $z=\pm L_z/2$, respectively. Because the electrode potentials are prescribed as spatially uniform Dirichlet conditions,
\[
\Phi(x,y,\pm L_z/2) = \pm \frac{V_{\text{applied}}}{2},
\]
their DCT representations are nonzero only for the zero transverse mode $(k_x,k_y)=(0,0)$. Specifically,
\begin{equation}
\hat{\Phi}_{\mathrm{el}}^{+}(k_x,k_y) =
\begin{cases}
+\dfrac{V}{2}, & (k_x,k_y)=(0,0),\\[6pt]
0, & \text{otherwise},
\end{cases}
\qquad
\hat{\Phi}_{\mathrm{el}}^{-}(k_x,k_y) =
\begin{cases}
-\dfrac{V}{2}, & (k_x,k_y)=(0,0),\\[6pt]
0, & \text{otherwise}.
\end{cases}
\label{eq:electrode_modes}
\end{equation}
All higher transverse modes are therefore driven solely by the transmembrane voltage $\hat{V}_m(k_x,k_y)$ through the interface jump condition and decay exponentially away from the membrane.
\subsubsection{Exact ionic current evaluation}
The electrolyte-to-membrane current density $J_{\text{elec}}$ is evaluated analytically in spectral space by differentiating the hyperbolic solution at the interface. This avoids the truncation errors associated with finite-difference approximations on the grid:
\begin{equation}
    \hat{J}_{\text{elec}} = \lambda k_{xy} \left[ \frac{\hat{\Phi}_{\text{el}}^{+} - \hat{\Phi}_{\text{el}}^{-}}{2 \sinh(k_{xy} L)} - \frac{\hat{V}_m}{2} \coth(k_{xy} L) \right].
\label{eq:J_spectral_final}
\end{equation}
For the zero mode ($k_{xy}=0$), the expression reduces to the classical linear form $J_{\text{elec}} = \lambda (V_{\text{applied}} - V_m) / L_z$. The physical-space current $J_{\text{elec}}(x,y)$ is then obtained via an inverse DCT to drive the evolution of the transmembrane voltage.
\begin{figure}[ht!]
\centering
\begin{tikzpicture}[x=1cm,y=1.5cm, font=\small]

\def\xL{-2.5} \def\xR{ 2.5}
\def\zTop{ 2.0} \def\zBot{-2.0} \def\zMid{ 0.0}

\fill[gray!10] (\xL,\zMid) rectangle (\xR,\zTop); 
\fill[gray!5] (\xL,\zBot) rectangle (\xR,\zMid); 

\draw[ultra thick, blue] (\xL,\zMid) -- (\xR,\zMid);
\node[anchor=east, blue] at (-0.45,\zMid+0.15) {Membrane};
\node[anchor=east, blue] at (-0.79,\zMid-0.15) {$z=0$};

\node at (0, 1.1) {Region 2 ($\Phi^+$)};
\node at (0, -1.1) {Region 1 ($\Phi^-$)};

\node[anchor=west] at (\xR+0.2, \zTop) {$\Phi = +V_{\text{applied}}/2$};
\node[anchor=west] at (\xR+0.2, \zBot) {$\Phi = -V_{\text{applied}}/2$};
\node[anchor=west] at (\xR+0.2, \zMid) {$\Phi^+ - \Phi^- = V_m$};

\draw[<-, thick, red] (0, -0.5) -- (0, -0.1);
\draw[<-, thick, red] (0, 0.1) -- (0, 0.5);
\node[red, anchor=west] at (0.1, 0.3) {$J_{\text{elec}}$};

\end{tikzpicture}
\caption{Schematic of the singular-interface electrostatic domain. The potential jump $V_m$ is enforced at $z=0$, while the current flux $J_{\text{elec}}$ remains continuous across the interface.}
\label{fig:electrostatics_singular}
\end{figure}

\subsection{Transmembrane Voltage Solver}
The transmembrane potential, $V_m(x,y,t)$, is governed by the equation presented in \eqref{eq:governing_vm}. A simple explicit time-stepping scheme for this equation is severely restricted by the fast electrical charging dynamics, requiring a prohibitively small time step $\Delta t$ for stability. To overcome this limitation, we employ a more stable semi-implicit time integration scheme based on the Backward Euler method.

The time derivative is discretized as $(V_m^{n+1} - V_m^n) / \Delta t$. The term on the right-hand side that leads to stiffness, the leakage conductance, is treated implicitly. The computationally expensive ionic current, $J_{\text{elec}}$, which depends on the 3D potential field, is treated explicitly to avoid a fully-coupled non-linear solve. This results in the following discretized equation:
\begin{equation}
    \frac{V_m^{n+1} - V_m^n}{\Delta t} = \frac{1}{C_m(\phi^n)} \left[ J_{\text{elec}}^{n} - G_m(\phi^n)V_m^{n+1} \right] 
    \label{eq:Vm_update}
\end{equation}
Rearranging this equation to isolate the unknown $V_m^{n+1}$ terms on the left-hand side transforms the update into a system of linear equations:
\begin{equation}
      V_m^{n+1} = \frac{C_m(\phi^n) V_m^n + \Delta t J_{\text{elec}}^n}{C_{m}(\phi^n) + \Delta t \, G_{\text{m}}(\phi^n)} 
    \label{eq:vm_implicit_system}
\end{equation}
Algorithm~\ref{alg:main_loop} summarizes the coupled electroporation time-stepping procedure, in which the electrostatic solve updates $V_m$ and the resulting electrical pressure feeds back into the phase-field evolution.

\begin{algorithm}[ht!]
\caption{Copuled electroporation algorithm}
\label{alg:main_loop}
\begin{algorithmic}[1]
\State Initialize phase field $\phi^0$ and transmembrane voltage $V_m^0$ (typically zero).
\For{$n = 0$ \textbf{to} $N_{\text{steps}}-1$}
    \State Compute pore/lipid mask $H^n \gets H(\phi^n)$ and blended maps $C_m^n,\ G_m^n$.
    \State Impose electrode BCs $\Phi=\pm V_{\text{applied}}/2$ and the membrane jump $\Phi(0^+)-\Phi(0^-)=V_m^n$ using Eq.~\eqref{eq:jump_jump}.

    \State Solve $\nabla^2\Phi^n=0$ using the spectral solver in Eqs.~\eqref{eq:hemlholtz_eq}--\eqref{eq:Phi_analytic}.
    \State Evaluate ionic current $J_{\mathrm{elec}}^n$ from one-sided $z^\pm$ gradients at the membrane in Eqs.~\eqref{eq:J_spectral_final}
    \State Update $V_m^{n+1}$ with the semi-implicit charging step in Eq.~\eqref{eq:vm_implicit_system}.
    \State Compute electrical pressure $P_{\mathrm{elec}}^n(V^{n+1}_m,H(\phi^n))$ in Eq.~\eqref{eq:p_elec_maxwell}.
    \State Set effective tension $\sigma_{\mathrm{eff}}^n \gets \sigma_0 + P_{\mathrm{elec}}^n$.
    \State Advance $\phi^{n+1}$ in Eq.~\eqref{eq:phi_hat}, adding thermal noise if enabled.
\EndFor
\end{algorithmic}
\end{algorithm}

\section{Results and Discussion}
In this section, we present a series of numerical experiments designed to demonstrate the versatility and robustness of our coupled diffuse-interface framework. Rather than restricting the model to a single parameter regime, we evaluate its performance across a diverse range of physical scales-from the nanosecond kinetics of pore nucleation to the microsecond dynamics of pore expansion. By varying key physical parameters such as line tension, surface tension, and phase-field mobility across these cases, we highlight the model's ability to faithfully capture the underlying electromechanical competition while maintaining numerical stability. This multi-scale validation underscores the utility of the method as a predictive tool for complex membrane permeabilization protocols.
\subsection{Verification of Critical Radius Dynamics}
With electrical terms disabled, single-pore simulations reproduce the critical-radius bifurcation reported for polymeric membranes, isolating line-tension vs. surface-tension mechanics~\cite{PhysRevLett.117.257801}.
 For a single circular pore of radius $R$, the free energy in Equation~(\ref{eq:sharp_interface}) reduces to 
 \begin{equation}
    {E}(R)=2\pi R \gamma - \pi R^2 \sigma, 
    \label{eq:single_pore_model}
 \end{equation}
where the first term penalizes creation of edge and the second term accounts for the reduction of membrane area under tension.  
Setting $dE/dR=0$ gives the critical radius $R_c=\gamma/\sigma$, which separates closing from growing pores.  
Using $\gamma = 1.5 \times 10^{-11}\,\mathrm{J\ m^{-1}}$ and $\sigma = 5 \times 10^{-4}\,\mathrm{J\ m^{-2}}$ as an example yields $R_c = 30\ \mathrm{nm}$.  

In the simulations, the instantaneous pore radius $R(t)$ is computed from the pore area $A_{\mathrm{pore}}(t)$:
\begin{equation}
R(t) = \sqrt{{A_{\mathrm{pore}}(t)}/{\pi}}, \quad A_{\mathrm{pore}}(t) = \sum_{i,j} \bigl[ 1 - H(\phi_{i,j}(t)) \bigr] \, h^2     
\end{equation}
where $H(\phi_{i,j})$ is the smooth step function of the phase field and $h=\Delta x=\Delta y=L_x/N_x$ for $L_x=L_y=1\ \mu\mathrm{m}$ and $N_x=N_y=256$. The interface thickness was set to $\varepsilon = 0.5\,h$ and $M=1\times 10^{6}\ \mathrm{m^{2}J^{-1}s^{-1}}$.

\begin{figure}[ht!]
\centering
\begin{subfigure}[c]{0.3\textwidth}
    \includegraphics[width=\textwidth]{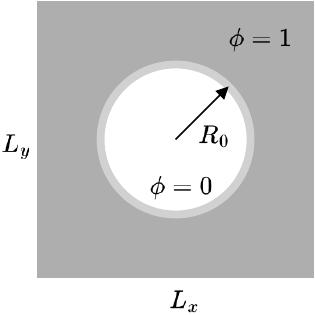}
    \caption{}
    \label{fig:domain}
\end{subfigure}%
\hspace{0.02\textwidth} 
\begin{subfigure}[c]{0.5\textwidth}
    \includegraphics[width=\textwidth]{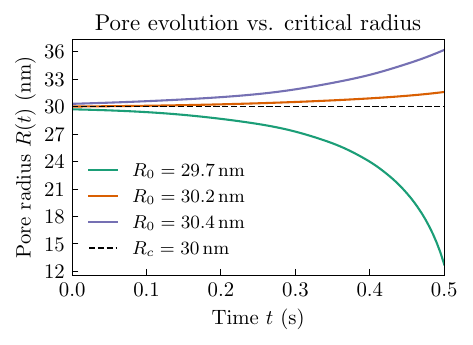}
    \caption{}
    \label{fig:results}
\end{subfigure}
\caption{(a) Model setup with a central pore of radius $R_0$ at the center (b) Pore-radius evolution: pores with $R_0<R_c$ close, whereas $R_0>R_c$ leads to expansion.}
\label{fig:validation}
\end{figure}

As shown in Fig.~\ref{fig:validation}, pores with $R_0 < R_c$ shrink and reseal, whereas those with $R_0 > R_c$ expand, reproducing the sharp-interface bifurcation behavior without parameter tuning and confirming the quantitative accuracy of the diffuse-interface formulation.

We performed a quantitative grid convergence study. The simulated critical radius, $R_c^{\text{sim}}$, was approximated on a series of progressively finer grids while keeping the interface-to-grid-spacing ratio constant at $\varepsilon/h=0.5$. The results, summarized in Table~\ref{tab:convergence}, show that $R_c^{\text{sim}}$ clearly converges to the theoretical value as the grid spacing $h$ is reduced.
\begin{table}[ht!]
\centering
\caption{Grid convergence study for the simulated critical radius, showing convergence to the theoretical value of $R_c = 30$ nm.}
\label{tab:convergence}
\begin{tabular}{c c c c c c}
\toprule
\textbf{Level} & \textbf{$N_x,\ N_y$} & \textbf{$h$ (nm)} & \textbf{ $\varepsilon$ (nm)} & \textbf{$R_c^{\text{sim}}$ (nm)} & \textbf{Error (\%)} \\
\midrule
1 & 64  & 15.87  & 7.94  & 32.84  & 9.4\% \\
2 & 128 & 7.87   & 3.94  & 30.60  &  2.0\% \\
3 & 256 & 3.92  & 1.96  &  30.15 & 0.4\% \\
4 & 512 & 1.95   & 0.98   & 30.08  & 0.3\% \\
\bottomrule
\end{tabular}
\end{table}

\subsection{Electric Field Focusing}

We quantify field focusing around a static, circular pore of radius $R_0=1.0~\mu\mathrm{m}$ in an otherwise intact membrane subject to an applied potential $V_\text{applied}=5.0~\mathrm{V}$. The electrolyte conductivity is $\lambda=1.0~\mathrm{S/m}$ with membrane thickness of $d_m=10\ \text{nm}$. Membrane (lipid) properties are $C_{\mathrm{lipid}}=0.01~\mathrm{F/m^2}$ and $G_{\mathrm{lipid}}=10^{-7}~\mathrm{S/m^2}$, while the pore is taken to be nearly non-capacitive and highly conductive ($G_{\mathrm{pore}} = 10^8~\mathrm{S/m^2}$). The computational domain is $L_x=L_y=10~\mu\mathrm{m}$ and $L_z=20~\mu\mathrm{m}$.

\begin{figure}[ht!]
\centering
\begin{subfigure}[c]{0.4\textwidth}
    \includegraphics[width=\textwidth]{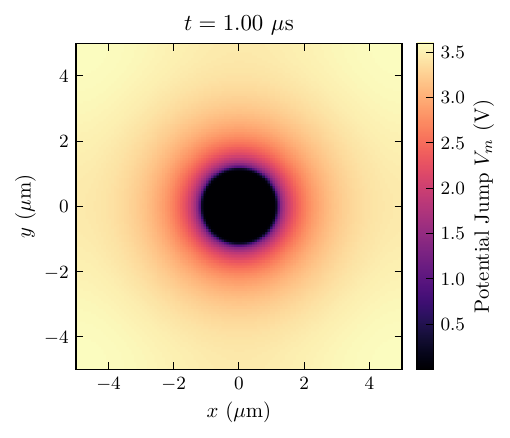}
    \caption{}
    \label{fig:vm_map} 
\end{subfigure}%
\hspace{0.02\textwidth}
\begin{subfigure}[c]{0.4\textwidth}
    \includegraphics[width=\textwidth]{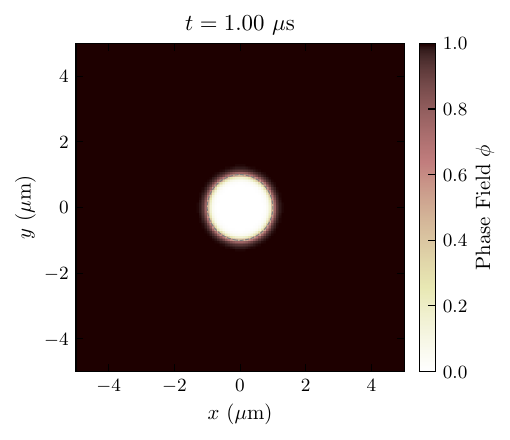}
    \caption{}
    \label{fig:pf_map}
\end{subfigure}
\hspace{0.02\textwidth}
\begin{subfigure}[c]{0.40\textwidth}
    \includegraphics[width=\textwidth]{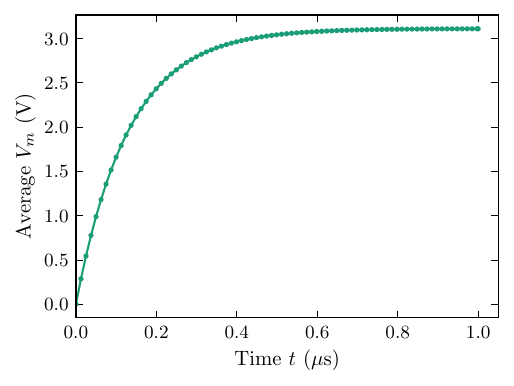}
    \caption{}
    \label{fig:charge_curve}
\end{subfigure}
\hspace{0.02\textwidth}
\begin{subfigure}[c]{0.4\textwidth}
    \includegraphics[width=\textwidth]{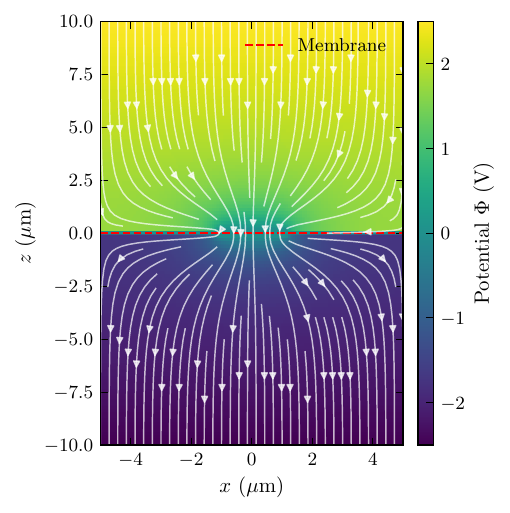}
    \caption{}
    \label{fig:pot_slice}
\end{subfigure}
\caption{(a) the membrane potential shows a clear depression in the vicinity of the pore; (b) the phase field delineates the pore geometry; (c) the area-averaged transmembrane voltage exhibits the expected charging dynamics; and (d) a cross-section reveals that the water-filled pore, being far more conductive than the lipid, acts as a low-resistance pathway-equipotential contours visibly converge and funnel through the pore.}
\label{fig:focusing} 
\end{figure}

Figure~\ref{fig:focusing} illustrates the multi-physics coupling. The potential field $\Phi$ exhibits characteristic "funneling" (Fig.~\ref{fig:focusing}d), where equipotential contours converge toward the conductive pore. This results in a localized depression of $V_m$ near the pore rim (Fig.~\ref{fig:focusing}a), a feature that deterministic models often miss but which is critical for accurate electromechanical stress calculation.

To verify the robustness of the semi-analytical approach, we conducted a grid-refinement study on the electrostatic solver by varying $N_z$ and keeping $N_x=N_y=128$. We recorded the steady-state area-averaged transmembrane voltage $\overline{V}_m$ and the total pore current 
\begin{equation}
    I_{\mathrm{pore}}=\int_{\Gamma_{\text{pore}}} J_{\mathrm{elec}}\,\mathrm{d}A.    
    \label{eq:pore_current}
\end{equation}
 As shown in Table~\ref{tab:grid_refinement_minimal}, the spectral solver exhibits remarkable accuracy and grid independence. Specifically, $\overline{V}_m$ remains constant across all levels, while $I_{\text{pore}}$ converges with a residual error of only $0.25\%$ at the finest resolution.
 \begin{table}[ht!]
\centering
\caption{Grid refinement study for the electrostatic solver.}
\label{tab:grid_refinement_minimal}
\begin{tabular}{cccccc}
\hline
\textbf{Level} & \textbf{ $N_z$} & \textbf{$\overline{V}_m$ (V)}  & \textbf{$I_{\text{pore}}$ ($\mu$A)} & \textbf{$|\Delta I_{\text{pore}}|$} \\
\hline
1 & 65   & 3.06 &  -5.18 &    - \\
2 & 129  & 3.06 &  -5.23 &   1.0\%  \\
3 & 257  & 3.06 &   -5.20 &  0.6\% \\
4 & 513  & 3.06 &   -5.17 &  0.6\% \\
\hline
\end{tabular}
\end{table}

\subsection{Pore Dynamics under Applied Voltage}
We investigate the electromechanical competition between line tension and electrical pressure by simulating the response of a single pre-existing pore to varying external potentials. The membrane, with an initial pore radius of $R_0 = 88$ nm, is modeled in a $1 \times 1 \times 2$ $\mu\text{m}^3$ domain discretized on a $128 \times 128 \times 129$ grid. Physical parameters are set to a line tension $\gamma = 1.5 \times 10^{-10}$ J m$^{-1}$ and a phase-field mobility $M = 5 \times 10^8$ m$^2$ J$^{-1}$ s$^{-1}$, with an interface thickness $\varepsilon = 1.5\Delta x$.

\begin{figure}[ht!]
    \centering
    \begin{subfigure}[b]{0.49\textwidth}
        \centering
        \includegraphics[width=\linewidth]{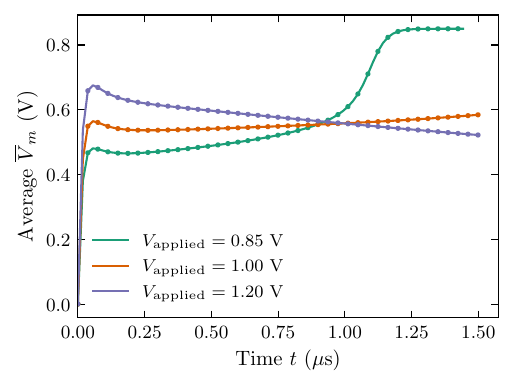}
        \caption{}
        \label{fig:voltage_comp}
    \end{subfigure}
    \vspace{0.3cm} 
    \begin{subfigure}[b]{0.49\textwidth}
        \centering
        \includegraphics[width=\linewidth]{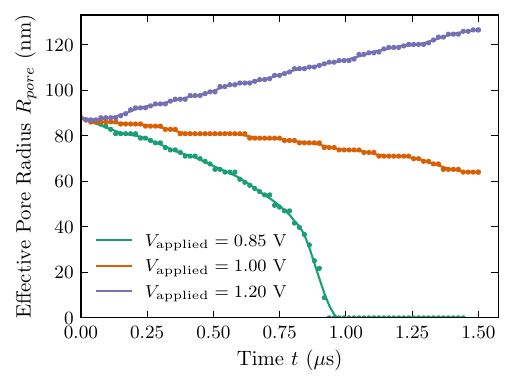}
        \caption{}
        \label{fig:radius_comp}
    \end{subfigure}

    \caption{Membrane dynamics for applied voltages near the threshold. (a) Average transmembrane potential $\overline{V}_m$ (charging curve) over time. (b) Pore radius $R(t)$ showing expansion above critical voltage and closure below the critical. }
    \label{fig:voltage_sweep}
\end{figure}

The temporal evolution of the system, illustrated in Fig.~\ref{fig:voltage_sweep}, reveals a clear bifurcation in pore dynamics governed by the applied voltage. At lower potentials (e.g., $V_\text{applied} = 0.85$ V), the restorative line tension dominates the Maxwell stress, forcing the pore to contract and eventually reseal. Conversely, exceeding a critical voltage threshold triggers expansion; as the membrane charges, the increasing electrical pressure overcomes the line tension penalty, driving the pore to grow toward the domain boundaries.

\begin{figure}[ht!]
    \centering
    
    \begin{minipage}[c]{0.03\textwidth}
        \rotatebox{90}{{0.85 V}}
    \end{minipage}
    \hspace{0.01\textwidth}
    \begin{subfigure}[c]{0.30\textwidth}
        \includegraphics[width=\linewidth]{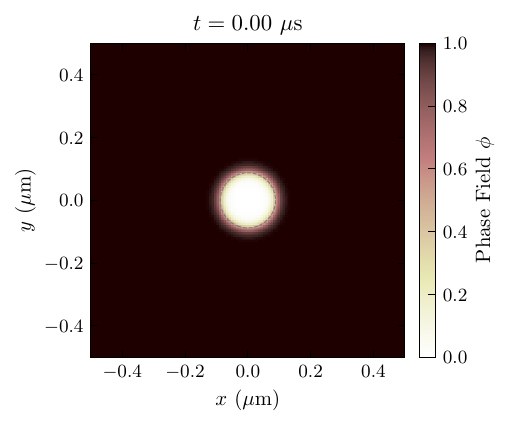}
    \end{subfigure}
    \hfill
    \begin{subfigure}[c]{0.30\textwidth}
        \includegraphics[width=\linewidth]{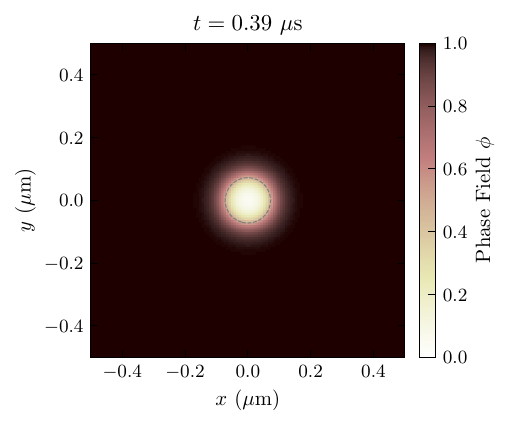}
    \end{subfigure}
    \hfill
    \begin{subfigure}[c]{0.30\textwidth}
        \includegraphics[width=\linewidth]{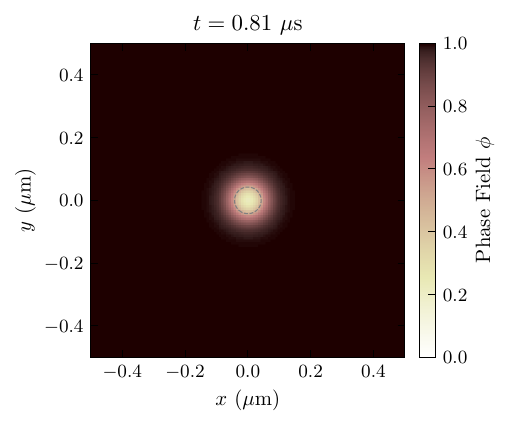}
    \end{subfigure}

    \vspace{0.3cm}

    \begin{minipage}[c]{0.03\textwidth}
        \rotatebox{90}{1.00 V}
    \end{minipage}
    \hspace{0.01\textwidth}
    \begin{subfigure}[c]{0.30\textwidth}
        \includegraphics[width=\linewidth]{pics/v_all_0.pdf}
    \end{subfigure}
    \hfill
    \begin{subfigure}[c]{0.30\textwidth}
        \includegraphics[width=\linewidth]{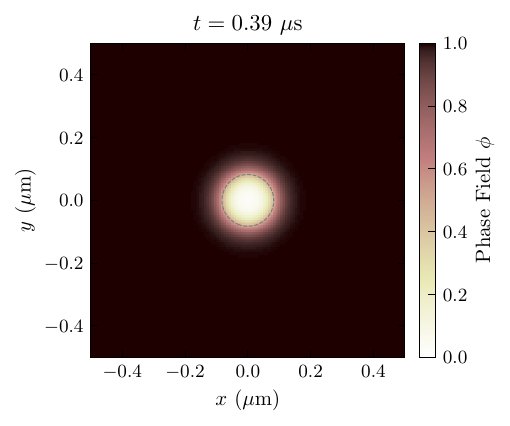}
    \end{subfigure}
    \hfill
    \begin{subfigure}[c]{0.30\textwidth}
        \includegraphics[width=\linewidth]{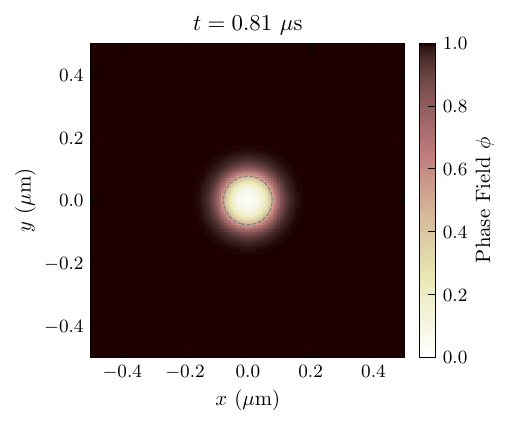}
    \end{subfigure}

    \vspace{0.3cm}

    \begin{minipage}[c]{0.03\textwidth}
        \rotatebox{90}{{1.20 V}}
    \end{minipage}
    \hspace{0.01\textwidth}
    \begin{subfigure}[c]{0.30\textwidth}
        \includegraphics[width=\linewidth]{pics/v_all_0.pdf}
    \end{subfigure}
    \hfill
    \begin{subfigure}[c]{0.30\textwidth}
        \includegraphics[width=\linewidth]{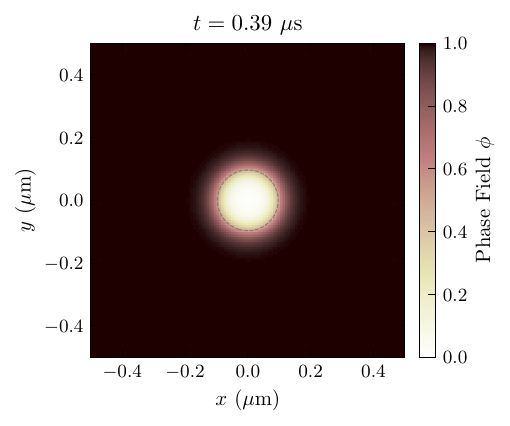}
    \end{subfigure}
    \hfill
    \begin{subfigure}[c]{0.30\textwidth}
        \includegraphics[width=\linewidth]{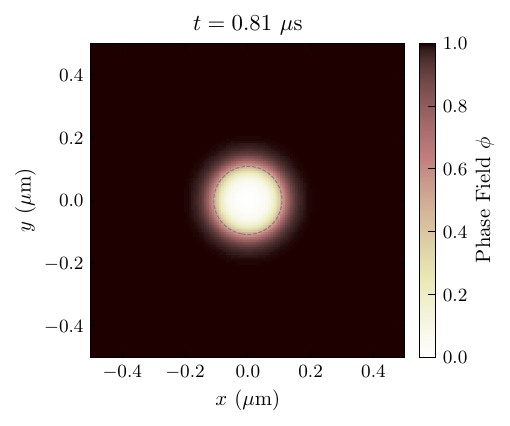}
    \end{subfigure}

    \caption{Parametric study of applied voltage influence on electroporation dynamics. Row labels indicate the constant applied potential. Columns (left to right): shows the pore size.}
    \label{fig:voltage_sweep_grid}
\end{figure}

The spatial manifestation of this behavior is captured in the snapshots provided in Fig.~\ref{fig:voltage_sweep_grid}. Each row illustrates the morphological evolution of the membrane at a specific voltage level. In the sub-critical regime ($0.85$ V), the phase-field contours show the gradual collapse of the initial defect until the membrane is fully restored. In contrast, at $1.20$ V, the model captures the rapid radial expansion of the pore, demonstrating how the electrical driving force scales with the applied potential to dictate the final state of the lipid bilayer.

\subsection{Spontaneous Electroporation}
To demonstrate spontaneous electroporation, we simulate a voltage-driven membrane starting from an intact configuration with no prescribed defects. Unlike deterministic simulations in previous examples, pore nucleation arises naturally from thermally induced fluctuations in the phase-field dynamics. 

Thermal effects are incorporated by augmenting the Allen--Cahn equation in (\ref{eq:AC_code}) with an additive stochastic forcing term,
\begin{equation}
\frac{\partial \phi}{\partial t} = - M \frac{\delta E}{\delta \phi} + \eta(x,y,t),
\end{equation}
where $\eta(x,y,t)$ is a Gaussian white noise process satisfying the fluctuation-dissipation theorem. In the numerical implementation, the noise is discretized as $A_{\text{noise}} = \sqrt{2 M k_B T / (\Delta x\Delta y \Delta t)}$ and localized to the lipid phase using a cubic smooth-step mask $H(\phi)$. Theretofore, the phase-field evolution in Eq.~\eqref{eq:phi_hat} becomes:
\begin{equation}
\hat{\phi}^{\,n+1}
=
\frac{
\hat{\phi}^{\,n}
+\Delta t\,\mathcal{F}\!\left[
-M\left(
\left(\frac{\gamma}{C_g}\right)\frac{g'(\phi^n)}{\varepsilon}
+\sigma_{\mathrm{tot}}^{\,n}\,H'(\phi^n)
\right)
+\eta^n(x,y)
\right]
}{
1+\Delta t\,M\left(\frac{\gamma}{C_g}\right)\varepsilon\,k^2
},
\label{eq:phi_hat_noise}
\end{equation}
where
\begin{equation}
\eta^n(x,y)=A_{\mathrm{noise}}\,\xi^n(x,y)\,H(\phi^n(x,y)),
\qquad
\xi^n \sim \mathcal{N}(0,1),
\label{eq:noise_mask}
\end{equation}

We first perform a simulation with nominal values and establish a baseline and then later we look into the impact of inputs in final electroporation voltage and effective pore radius.

\subsection{Numerical Regularization in Spontaneous Electroporation}
While the spectral evaluation of $J_{\text{elec}}$ in Eq.~(\ref{eq:J_spectral_final}) provides high-order accuracy for smooth potential distributions, the onset of stochastic nucleation introduces sharp local gradients and high-frequency content. The spectral operator for the ionic current involves a $k_{xy} \coth(k_{xy} L)$ multiplier, which effectively acts as a high-pass filter. In the presence of the white noise term $\eta(x,y,t)$, this operator tends to amplify grid-scale fluctuations, which can lead to unphysical ``ringing'' artifacts or Gibbs-like oscillations at the nascent pore rim.

To maintain numerical stability during these emergent events, we employ a hybrid numerical strategy. The electric potential $\Phi$ is solved using the semi-analytical spectral method to accurately capture global field focusing. However, the interface gradient $\partial_z \Phi$ used to calculate $J_{\text{elec}}$ is evaluated using a second-order, one-sided finite difference (FD) approximation in physical space. Specifically, let $k_m$ denote the grid index of the membrane interface at $z=0$, and $\Delta z$ be the vertical grid spacing. The gradients from the upper ($+$) and lower ($-$) electrolyte regions are computed as:
\begin{align}
    \left. \partial_z \Phi^+ \right|_{z=0^+} &\approx \frac{-3\Phi_{k_m} + 4\Phi_{k_m+1} - \Phi_{k_m+2}}{2\Delta z}, \\
    \left. \partial_z \Phi^- \right|_{z=0^-} &\approx \frac{3\Phi_{k_m} - 4\Phi_{k_m-1} + \Phi_{k_m-2}}{2\Delta z}.
\end{align}
The effective ionic current density is then averaged across the interface to enforce flux continuity:
\begin{equation}
    J_{\text{elec}} = \frac{1}{2} \lambda \left( \left. \partial_z \Phi^+ \right|_{z=0^+} + \left. \partial_z \Phi^- \right|_{z=0^-} \right).
\end{equation}

This choice provides a localized numerical regularization at the $2\Delta z$ scale, damping sub-grid oscillations that are physically irrelevant to the continuum phase-field evolution but numerically destabilizing in a purely spectral context. This hybrid approach ensures robust convergence of the transmembrane voltage $V_m$ even as the membrane topology undergoes rapid, noise-driven transitions.

\subsubsection{Baseline Simulation}
The simulation is performed on a $L_x=L_y=100~\text{nm}$ and $L_z= 20~\text{$\mu$m}$ domain with $N_x=N_y=128,\ N_z=129$ at $T = 310~\text{K}$. The membrane is characterized by a mobility $M = 5.0 \times 10^7~\text{m}^{2}\text{J}^{-1}\text{s}^{-1}$ and a line tension $\gamma = 1.5 \times 10^{-11}~\text{J}\text{m}^{-1}$. The applied potential is set to $V_{\text{applied}} = 1.5~\text{V}$, providing a strong electrical driving force that lowers the nucleation barrier.
\begin{figure}[ht!]
\centering
\begin{subfigure}[c]{0.44\textwidth}
    \includegraphics[width=\textwidth]{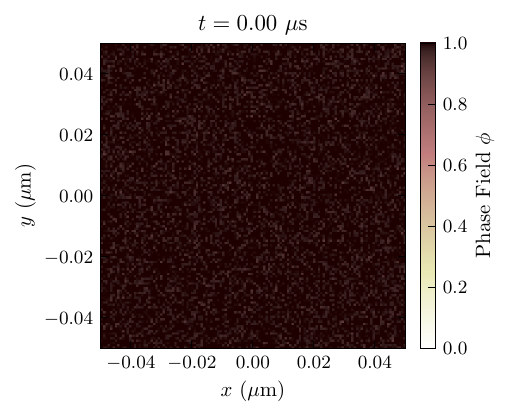}
    \caption{}
    \label{fig:electroporation_0} 
\end{subfigure}%
\hspace{0.02\textwidth}
\begin{subfigure}[c]{0.44\textwidth}
    \includegraphics[width=\textwidth]{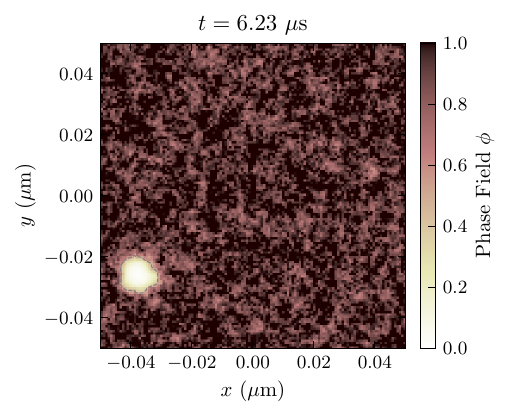}
    \caption{}
    \label{fig:electroporation_1}
\end{subfigure}
\hspace{0.02\textwidth}
\begin{subfigure}[c]{0.44\textwidth}
    \includegraphics[width=\textwidth]{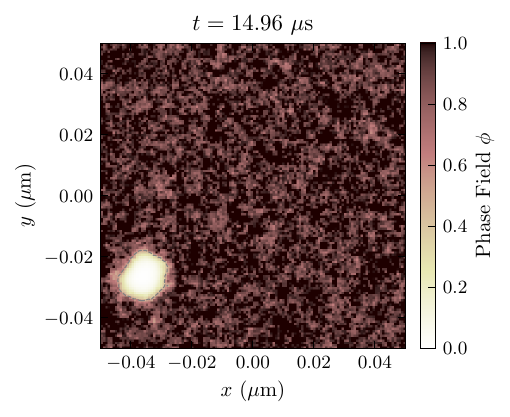}
    \caption{}
    \label{fig:electroporation_2}
\end{subfigure}
\hspace{0.02\textwidth}
\begin{subfigure}[c]{0.44\textwidth}
    \includegraphics[width=\textwidth]{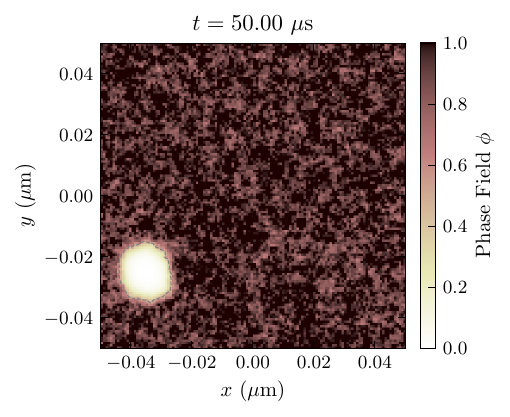}
    \caption{}
    \label{fig:electroporation_3}
\end{subfigure}
\caption{From (a) to (d), the evolution shows a pore nucleating, growing, and then slowing down as it reaches an almost stabilized size.}
\label{fig:stochastic_nucleation}
\end{figure}

The simulation captures the rapid interplay between capacitive charging and stochastic pore nucleation. Initially, the transmembrane voltage $V_m$ rises toward the applied potential as the lipid bilayer charges. Once $V_m$ reaches the breakdown threshold, thermal fluctuations trigger the emergence of transient defects (Fig.~\ref{fig:stochastic_nucleation}b). 

A competitive nucleation process follows: as soon as a single defect exceeds the critical radius and stabilizes into a hydrophilic pore, it acts as a low-resistance shunt. This leads to an almost immediate discharge of the membrane, evidenced by a sharp drop in $V_m$. This discharge localized to the pore reduces the electrical pressure across the remaining membrane, causing neighboring sub-critical defects to reseal under the restorative influence of line tension (Fig.~\ref{fig:stochastic_nucleation}c). The sensitivity of this breakdown event to various physical inputs is explored in the following parametric studies.
\begin{figure}[ht!]
\centering
\begin{subfigure}[c]{0.48\textwidth}
    \includegraphics[width=\textwidth]{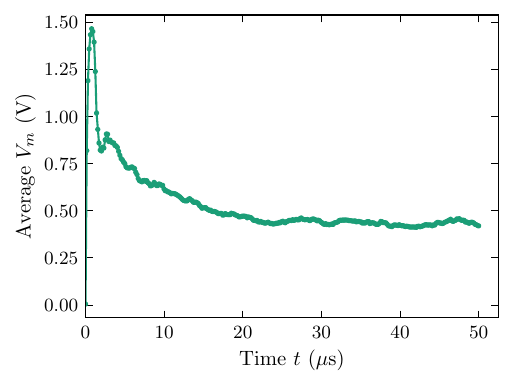}
    \caption{}
    \label{fig:electropore_vm_history} 
\end{subfigure}
\hfill
\begin{subfigure}[c]{0.48\textwidth}
    \includegraphics[width=\textwidth]{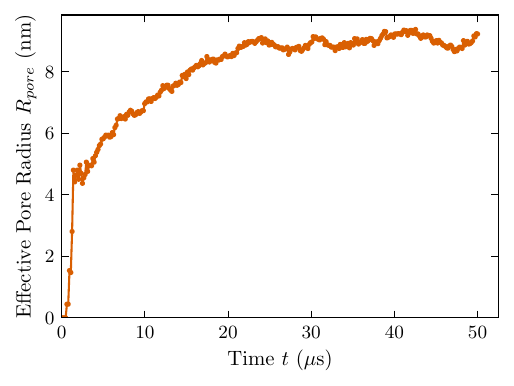}
    \caption{}
    \label{fig:electropore_pf_history}
\end{subfigure}
\caption{(a) Average transmembrane voltage $\overline{V}_m$ showing the initial capacitive charging phase followed by a rapid discharge upon membrane breakdown. (b) Evolution of the effective pore radius, illustrating the transition from stochastic nucleation to the growth and stabilization of hydrophilic pore.}
\label{fig:vm_and_pore_radius}
\end{figure}

This competition is quantitatively visible in Fig.~\ref{fig:vm_and_pore_radius}. 

\subsubsection{Effect of Applied Voltage}
\begin{figure}[ht!]
\centering
\begin{subfigure}[c]{0.48\textwidth}
    \includegraphics[width=\textwidth]{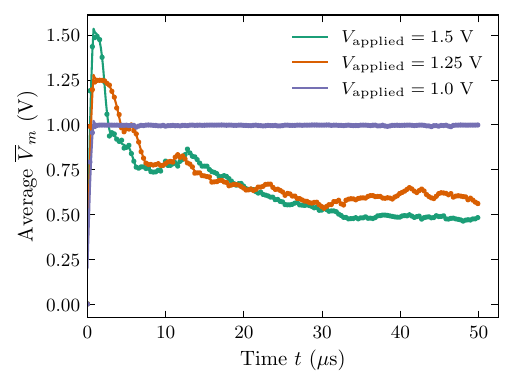}
    \caption{}
    \label{fig:electropore_vm_map_history_V} 
\end{subfigure}
\hfill
\begin{subfigure}[c]{0.48\textwidth}
    \includegraphics[width=\textwidth]{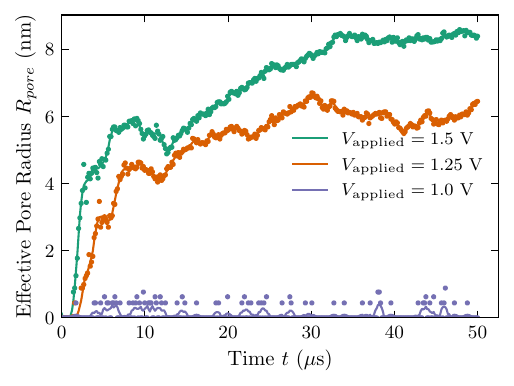}
    \caption{}
    \label{fig:electropore_pf_map_history_V}
\end{subfigure}
\caption{Influence of applied voltage $V_{\text{applied}}$ on nucleation and growth. Higher potentials accelerate membrane breakdown, leading to larger equilibrium pores and rapid capacitive discharge. At $V_{\text{applied}} = 1$~V, the system remains in a sub-critical regime characterized by the formation of transient, flickering defects.}
\label{fig:vm_and_pore_radius_V}
\end{figure}

We first investigate the role of the applied potential $V_{\text{applied}}$ in electroporation, while keeping the mobility fixed at $M = 5\times10^{7}~\mathrm{m}^{2}\,\mathrm{J}^{-1}\,\mathrm{s}^{-1}$. As shown in Fig.~\ref{fig:vm_and_pore_radius_V}, the applied voltage serves as the primary control parameter governing pore nucleation.

At higher applied voltages, the system exhibits the formation of stable pores. This behavior arises because the steady-state balance between the external power supply and the pore leakage current shifts toward larger conductive areas in order to accommodate the increased potential. In particular, applied voltages of $1.25$ and $1.5~\mathrm{V}$ lead to the emergence of stable pores that grow gradually until reaching an equilibrium radius. Concurrently, the transmembrane voltage $V_m$ relaxes to a steady-state value.

By contrast, for an applied voltage of $1.0~\mathrm{V}$, the membrane remains fully charged with $V_m \approx 1~\mathrm{V}$ but does not sustain stable pores. Instead, only transient nanoscale pores ($<1~\mathrm{nm}$) intermittently form and collapse, indicating subcritical electroporation dynamics.

\subsubsection{Effect of Line Tension}
\begin{figure}[ht!]
\centering
\begin{subfigure}[c]{0.48\textwidth}
    \includegraphics[width=\textwidth]{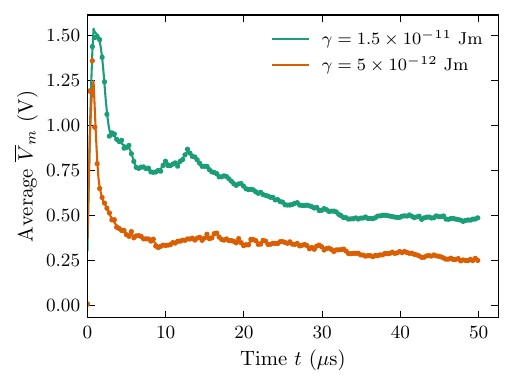}
    \caption{}
    \label{fig:electropore_vm_map_history_gamma} 
\end{subfigure}
\hfill
\begin{subfigure}[c]{0.48\textwidth}
    \includegraphics[width=\textwidth]{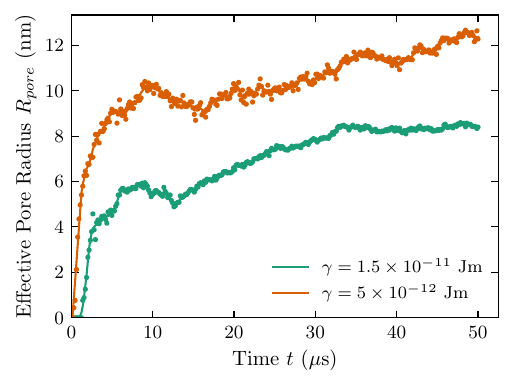}
    \caption{}
    \label{fig:electropore_pf_map_history_gamma}
\end{subfigure}
\caption{Impact of line tension $\gamma$ on poration kinetics. As the primary restorative force, higher line tension increases the energetic penalty for pore formation, resulting in smaller equilibrium radii and a reduced leakage current, which maintains a higher average transmembrane voltage.}
\label{fig:vm_and_pore_radius_gamma}
\end{figure}

Figure~\ref{fig:vm_and_pore_radius_gamma} illustrates the influence of the pore line tension $\gamma$ on the electroporation dynamics. While the applied voltage and membrane capacitance are identical across all cases, variations in $\gamma$ markedly alter the onset and growth of conductive defects. Lower line tension reduces the energetic penalty associated with pore perimeter, thereby lowering the effective nucleation barrier and promoting rapid pore expansion once a defect forms. This is reflected in the abrupt collapse of the transmembrane voltage $V_m$ and the rapid increase in the effective pore radius.

Conversely, increasing $\gamma$ stabilizes small defects and delays pore growth, resulting in a more gradual discharge of the membrane voltage and a suppressed pore radius over the same time interval. Importantly, the peak value of $V_m$ prior to breakdown remains nearly unchanged, indicating that line tension primarily governs the post-nucleation kinetics rather than the thermodynamic threshold for electroporation. These results highlight the critical role of line tension in controlling the temporal characteristics of membrane discharge and pore evolution.

\subsubsection{Effect of Surface Tension}
\begin{figure}[ht!]
\centering
\begin{subfigure}[c]{0.48\textwidth}
    \includegraphics[width=\textwidth]{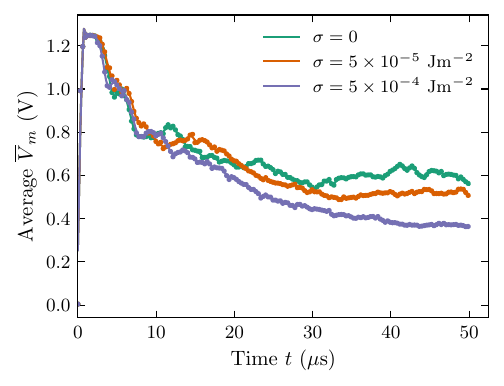}
    \caption{}
    \label{fig:electropore_vm_map_history_sigma} 
\end{subfigure}
\hfill
\begin{subfigure}[c]{0.48\textwidth}
    \includegraphics[width=\textwidth]{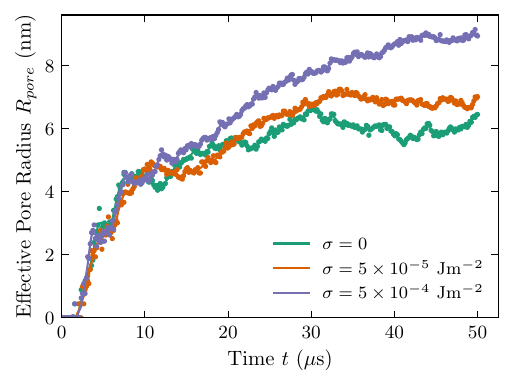}
    \caption{}
    \label{fig:electropore_pf_map_history_sigma}
\end{subfigure}
\caption{Influence of intrinsic surface tension $\sigma$ on membrane breakdown. Surface tension acts as a driving force for pore expansion, opposing the restorative effect of line tension. Higher values of $\sigma$ facilitate larger equilibrium pore radii and accelerate the subsequent discharge of the transmembrane potential.}
\label{fig:vm_and_pore_radius_sigma}
\end{figure}

We investigate the sensitivity of the nucleation process to the base membrane tension $\sigma$. As shown in Fig.~\ref{fig:vm_and_pore_radius_sigma}, higher initial tension results in the larger pore size. Conversely, at lower tensions, the membrane can sustain a higher $V_m$ before the stochastic fluctuations overcome the restorative line tension, leading to a more delayed discharge phase.

\subsubsection{Effect of Mobility}
\begin{figure}[ht!]
\centering
\begin{subfigure}[c]{0.48\textwidth}
    \includegraphics[width=\textwidth]{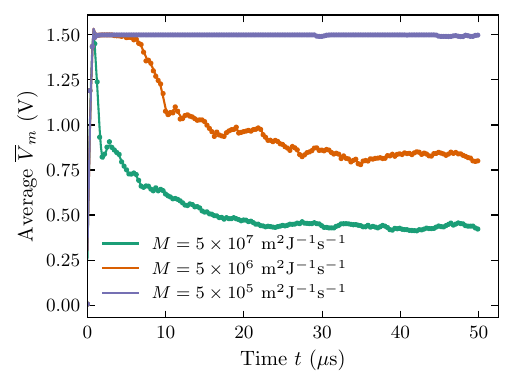}
    \caption{}
    \label{fig:electropore_vm_map_history_M} 
\end{subfigure}
\hfill
\begin{subfigure}[c]{0.48\textwidth}
    \includegraphics[width=\textwidth]{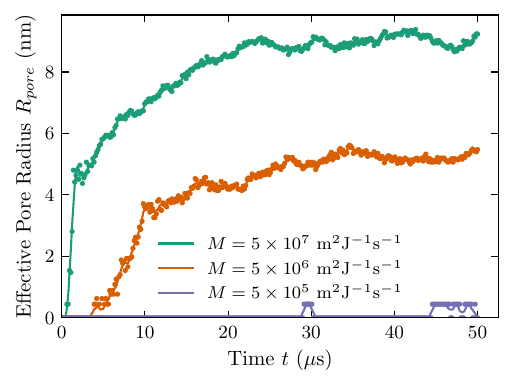}
    \caption{}
    \label{fig:electropore_pf_map_history_M}
\end{subfigure}
\caption{Influence of intrinsic surface tension $\sigma$ on membrane breakdown. Surface tension acts as a driving force for pore expansion, opposing the restorative effect of line tension. Higher values of $\sigma$ facilitate larger equilibrium pore radii and accelerate the subsequent discharge of the transmembrane potential.}
\label{fig:vm_and_pore_radius_M}
\end{figure}

The phase-field mobility $M$ controls the rate of lipid rearrangement within the membrane and may be interpreted as an effective inverse resistance to interfacial motion. In this sense, higher mobility corresponds to a more fluid-like membrane response, while lower mobility reflects increased resistance to pore growth. As illustrated in Fig.~\ref{fig:vm_and_pore_radius_M}, variations in $M$ do not significantly modify the thermodynamic breakdown voltage; instead, they strongly influence the \emph{kinetics} of membrane discharge following pore nucleation.

At higher mobility values, pores expand rapidly once nucleated, resulting in an abrupt increase in membrane conductance and a near-instantaneous collapse of the transmembrane voltage $V_m$. Conversely, lower mobility leads to a more gradual or ``soft'' breakdown, in which the transmembrane voltage remains elevated for an extended period as pore expansion proceeds slowly and competes with the electrolyte charging current. For sufficiently low mobility, specifically $M = 5\times10^{5}~\mathrm{m}^{2}\,\mathrm{J}^{-1}\,\mathrm{s}^{-1}$, pore growth is strongly suppressed and electroporation is either significantly delayed or does not occur within the simulated time window.

\section{Conclusion}
We presented a stabilized diffuse-interface continuum model for membrane electroporation that couples (i) Allen--Cahn pore evolution, (ii) a spatially varying leaky-dielectric description of the transmembrane voltage, and (iii) a quasi-static electrolyte potential governed by Laplace's equation. The central numerical contribution is a semi-implicit update for the transmembrane voltage in which the stiff leakage term is treated implicitly while the electrolyte-to-membrane ionic current is lagged. This yields a closed-form local update for $V_m$ that removes the severe timestep restriction associated with the fast dielectric relaxation time and enables robust simulations across a wide range of spatial resolutions.

To compute the electrolyte potential efficiently and accurately, we introduced a semi-analytical spectral Laplace solver that applies a two-dimensional DCT in the membrane plane and reduces the three-dimensional problem to independent one-dimensional Helmholtz problems in the electrode-normal direction. The resulting solutions are available in closed form for each transverse mode, which enables an exact evaluation of the ionic current at the membrane and avoids finite-difference truncation errors at the interface. In combination with a non-local Maxwell-stress-based electrical pressure that remains well-defined as pores become conductive, the coupled method remains stable under grid refinement and captures key electroporation hallmarks, including the sharp-interface critical-radius bifurcation and strong electric-field focusing through conductive defects.

Finally, by augmenting the phase-field dynamics with thermal noise, the framework reproduces stochastic pore nucleation and emergent breakdown events without prescribing initial defects, and it correctly reflects how applied voltage, line tension, surface tension, and mobility shape both the thermodynamic threshold and the kinetics of discharge. Overall, the proposed formulation provides a compact and computationally efficient continuum tool for predictive studies of electroporation-driven permeabilization and for the design of electrically controlled release protocols. Future work will incorporate curvature elasticity, spatially heterogeneous membrane properties, and electrolyte models beyond Laplace electrostatics to address regimes where ion transport, concentration polarization, or membrane deformation become essential.


\section*{Declaration of Competing Interest}
The author declare that they have no known competing financial interests or personal relationships that could have appeared to influence the work reported in this paper.

\bibliography{ref}

\end{document}